\documentclass[%
preprint,
nofootinbib,
 amsmath,amssymb,
 aps,
tightenlines]{revtex4-1}

\usepackage{color}
\usepackage{graphicx}% Include figure files
\usepackage{dcolumn}% Align table columns on decimal point
\usepackage{bm}% bold math
\begin{document}

%
% Bibliography and bibfile
\def\aj{AJ}%
          % Astronomical Journal
\def\actaa{Acta Astron.}%
          % Acta Astronomica
\def\araa{ARA\&A}%
          % Annual Review of Astron and Astrophys
\def\apj{ApJ}%
          % Astrophysical Journal
\def\apjl{ApJ}%
          % Astrophysical Journal, Letters
\def\apjs{ApJS}%
          % Astrophysical Journal, Supplement
\def\ao{Appl.~Opt.}%
          % Applied Optics
\def\apss{Ap\&SS}%
          % Astrophysics and Space Science
\def\aap{A\&A}%
          % Astronomy and Astrophysics
\def\aapr{A\&A~Rev.}%
          % Astronomy and Astrophysics Reviews
\def\aaps{A\&AS}%
          % Astronomy and Astrophysics, Supplement
\def\azh{AZh}%
          % Astronomicheskii Zhurnal
\def\baas{BAAS}%
          % Bulletin of the AAS
\def\bac{Bull. astr. Inst. Czechosl.}%
          % Bulletin of the Astronomical Institutes of Czechoslovakia 
\def\caa{Chinese Astron. Astrophys.}%
          % Chinese Astronomy and Astrophysics
\def\cjaa{Chinese J. Astron. Astrophys.}%
          % Chinese Journal of Astronomy and Astrophysics
\def\icarus{Icarus}%
          % Icarus
\def\jcap{J. Cosmology Astropart. Phys.}%
          % Journal of Cosmology and Astroparticle Physics
\def\jrasc{JRASC}%
          % Journal of the RAS of Canada
\def\mnras{MNRAS}%
          % Monthly Notices of the RAS
\def\memras{MmRAS}%
          % Memoirs of the RAS
\def\na{New A}%
          % New Astronomy
\def\nar{New A Rev.}%
          % New Astronomy Review
\def\pasa{PASA}%
          % Publications of the Astron. Soc. of Australia
\def\pra{Phys.~Rev.~A}%
          % Physical Review A: General Physics
\def\prb{Phys.~Rev.~B}%
          % Physical Review B: Solid State
\def\prc{Phys.~Rev.~C}%
          % Physical Review C
\def\prd{Phys.~Rev.~D}%
          % Physical Review D
\def\pre{Phys.~Rev.~E}%
          % Physical Review E
\def\prl{Phys.~Rev.~Lett.}%
          % Physical Review Letters
\def\pasp{PASP}%
          % Publications of the ASP
\def\pasj{PASJ}%
          % Publications of the ASJ
\def\qjras{QJRAS}%
          % Quarterly Journal of the RAS
\def\rmxaa{Rev. Mexicana Astron. Astrofis.}%
          % Revista Mexicana de Astronomia y Astrofisica
\def\skytel{S\&T}%
          % Sky and Telescope
\def\solphys{Sol.~Phys.}%
          % Solar Physics
\def\sovast{Soviet~Ast.}%
          % Soviet Astronomy
\def\ssr{Space~Sci.~Rev.}%
          % Space Science Reviews
\def\zap{ZAp}%
          % Zeitschrift fuer Astrophysik
\def\nat{Nature}%
          % Nature
\def\iaucirc{IAU~Circ.}%
          % IAU Cirulars
\def\aplett{Astrophys.~Lett.}%
          % Astrophysics Letters
\def\apspr{Astrophys.~Space~Phys.~Res.}%
          % Astrophysics Space Physics Research
\def\bain{Bull.~Astron.~Inst.~Netherlands}%
          % Bulletin Astronomical Institute of the Netherlands
\def\fcp{Fund.~Cosmic~Phys.}%
          % Fundamental Cosmic Physics
\def\gca{Geochim.~Cosmochim.~Acta}%
          % Geochimica Cosmochimica Acta
\def\grl{Geophys.~Res.~Lett.}%
          % Geophysics Research Letters
\def\jcp{J.~Chem.~Phys.}%
          % Journal of Chemical Physics
\def\jgr{J.~Geophys.~Res.}%
          % Journal of Geophysics Research
\def\jqsrt{J.~Quant.~Spec.~Radiat.~Transf.}%
          % Journal of Quantitiative Spectroscopy and Radiative Trasfer
\def\memsai{Mem.~Soc.~Astron.~Italiana}%
          % Mem. Societa Astronomica Italiana
\def\nphysa{Nucl.~Phys.~A}%
          % Nuclear Physics A
\def\physrep{Phys.~Rep.}%
          % Physics Reports
\def\physscr{Phys.~Scr}%
          % Physica Scripta
\def\planss{Planet.~Space~Sci.}%
          % Planetary Space Science
\def\procspie{Proc.~SPIE}%
          % Proceedings of the SPIE

%\preprint{APS/123-QED}

\title{Effects of electrically charged dark matter \\ on cosmic microwave background anisotropies}% Force line breaks with \\

\author{Ayuki Kamada}
 \email{ayuki.kamada@ucr.edu}
 \affiliation{%
 Department of Physics and Astronomy, University of California, Riverside, CA 92521, USA
}%
\author{Kazunori Kohri}
\email{kohri@post.kek.jp}
 \affiliation{%
 Institute of Particle and Nuclear Studies, KEK, 1-1 Oho, Tsukuba, Ibaraki 305-0801, Japan
}%
 \affiliation{%
 The Graduate University for Advanced Studies (SOKENDAI), 1-1 Oho, Tsukuba, Ibaraki 305-0801, Japan
}%
\author{Tomo Takahashi}
 \email{tomot@cc.saga-u.ac.jp}
 \affiliation{%
 Department of Physics, Saga University, Saga 840-8502, Japan
}%
\author{Naoki Yoshida}
\email{naoki.yoshida@phys.s.u-tokyo.ac.jp}
 \affiliation{%
 Kavli Institute for the Physics and Mathematics of the Universe (WPI), University of Tokyo Institutes for Advanced Study, University of Tokyo, Kashiwa 277-8583, Japan
}%
 \affiliation{%
 Department of Physics, University of Tokyo, Tokyo 113-0033, Japan
}%
 \affiliation{%
 CREST, Japan Science and Technology Agency, 4-1-8 Honcho, Kawaguchi, Saitama, 332-0012, Japan
}%


\date{\today}% It is always \today, today,

\begin{abstract}
We examine the possibility that dark matter consists of charged massive particles (CHAMPs) in view of the cosmic microwave background (CMB) anisotropies.
The evolution of cosmological perturbations of CHAMP with other components is followed in a self-consistent manner, without assuming that CHAMP and baryons are tightly coupled.
We incorporate for the first time the ``kinetic re-coupling" of the Coulomb scattering, which is characteristic of heavy CHAMPs.
By a direct comparison of the predicted CMB temperature/polarization auto-correlations in CHAMP models and the observed spectra in the {\it Planck} mission,
we show that CHAMPs leave sizable effects on them if they are lighter than $10^{11}$\,GeV.
Our result can be applicable to any CHAMP as long as its lifetime is much longer than the cosmic time at the recombination ($\sim 4 \times 10^{5}$\,yr).
An application to millicharged particles is also discussed.
\end{abstract}

\maketitle

%\tableofcontents

\section{Introduction}
\label{sec:intro}
It appears unreasonable at the first look that dark matter (DM) consists of charged massive particles (CHAMPs).
The possibility has not been completely ruled out after intensive works\,\cite{DeRujula:1989fe, Dimopoulos:1989hk, Gould:1989gw, McDermott:2010pa}.
If CHAMP is (quasi-)stable over the age of the Universe ($\tau_{\rm Ch} \gtrsim 10^{10}$\,yr), the stringent constraints are derived from multiple terrestrial efforts. 
One is a search of heavy isotopes in deep sea water (see\,\cite{Dimopoulos:1989hk, Agashe:2014kda} and references therein), 
from which one obtains that the mass of CHAMP is required to be $m_{\rm Ch} \gtrsim 10^{8}$\,GeV provided the relic density of CHAMP accounts for that of DM: $\Omega_{\rm Ch} \simeq \Omega_{\rm DM}$. 
Others seek energy deposition in detectors from DM scatterings.
Even a conservative lower bound on the CHAMP mass is as strong as $m_{\rm Ch} \gtrsim 10^{11}$\,GeV not to leave a significant number of events in cosmic-ray detectors (see\,\cite{Dimopoulos:1989hk} and references therein).

Such terrestrial experiments implicitly assume that DM particles are incoming to us with an expected rate.
This assumption, however, may be vulnerable to the existence of the Galactic magnetic field\,\cite{Chuzhoy:2008zy}.
It may expel CHAMPs from the the Galactic disk via the Fermi acceleration mechanism and prohibit CHAMPs from re-entering the Galactic disk due to a small gyroradius, opening up a CHAMP mass window of $10^5\,{\rm GeV} \lesssim m_{\rm Ch} \lesssim 10^{11}$\,GeV.
The same consideration in galaxy clusters potentially provide a stringent lower bound on the CHAMP mass ($m_{\rm Ch} \gtrsim 10^{14}$\,GeV) not to smear the DM density profile \cite{Kadota:2016tqq}.

The thermal relic abundance of heavy CHAMP ($m_{\rm Ch} \gtrsim 2 \times 10^5$\,GeV) is much larger than the critical density and thus overcloses the Universe\,\cite{DeRujula:1989fe, Dimopoulos:1989hk, Griest:1989wd}.
However, they are still interesting once nonthermal production mechanisms are addressed (see, e.g., \,\cite{Kudo:2001ie}).
This is reasonable because heavy CHAMPs can be thermally produced only if the reheating temperature of the Universe is as high as $T_{\rm RH} \gtrsim m_{\rm Ch}$.

If CHAMP decays into a neutral particle that accounts for DM at present, the stringent terrestrial constraints are not applicable.
On the other hand, decaying CHAMPs are still constrained by the catalyzed big bang nucleosynthesis (CBBN)\,\cite{Pospelov:2006sc, Kohri:2006cn, Steffen:2006hw, Hamaguchi:2007mp, Bird:2007ge, Kawasaki:2007xb, Pradler:2007is, Pospelov:2007js, Kawasaki:2008qe, Jittoh:2008eq, Pospelov:2008ta} and the suppression of structure formation\,\cite{Sigurdson:2003vy, Profumo:2004qt, Kohri:2009mi, Kamada:2013sh}, depending on their lifetimes.

In this paper, we consider that DM consists solely of heavy CHAMPs ($m_{\rm Ch} \gtrsim 10^{8}$\,GeV) at shortest until the recombination of the Universe.
Thus, our results can be applicable as long as the lifetime of CHAMP is longer than the cosmic time of the recombination ($\tau_{\rm Ch}\gtrsim 4 \times 10^{5}$\,yr), while the terrestrial experiments assume $\tau_{\rm Ch} \gtrsim 10^{10}$\,yr.
The electric charge of CHAMP is assumed to be the same as proton ($p^{+}$).
This is because even if elementary CHAMP is negatively charged, it forms a positively charged bound state with a He$^{2+}$ well before the recombination, $T \sim 10$\,keV\,\cite{Kohri:2009mi}.
Positively charged CHAMP becomes neutral finally by forming a bound state with an electron ($e^{-}$) at the recombination like proton.
Here is a difference if we consider millicharged particles (see \,\cite{Vinyoles:2015khy} for references). 
Millicharged particles are not neutralized at any time.
However, our discussion and calculation for CHAMP are not changed essentially even for millicharged particles as we will see below.

The Compton scattering with photon is subdominant because the cross section is suppressed by $1/m_{\rm Ch}^2$.
Before the recombination, however, CHAMPs are subject to the photon pressure through the Coulomb scattering with baryons.
The photon pressure disturbs CHAMP falling into the primordial gravitational potential.
The gravitational potentials undergo damped oscillations in CHAMP models unlike the standard cold dark matter (CDM) model.
There have been some analytical estimations of effects of CHAMP on the cosmic microwave background (CMB) anisotropies\,\cite{DeRujula:1989fe, Burrage:2009yz, McDermott:2010pa}. 
In this paper, we examine effects of the DM charge on the CMB anisotropies, especially temperature/polarization auto-correlations ($TT/EE$-power spectra) numerically.
By calculating the $TT/EE$-power spectra, we find that one can infer how stringent constraint on the CHAMP mass can be derived, in principle, from the precise measurements of the CMB anisotropies.

Our treatment of the collision term is based on the recent development (paper I \cite{Binder:2016pnr}).
As a consequence of the long-range nature of the Coulomb force, the interaction rate per Hubble time increases with the cosmic time to become unity before the recombination depending on the CHAMP mass.
We will call this ``kinetic re-coupling" in contrast to the usual kinetic decoupling of DM. 
It is not clear that the recent intensive works\,\cite{Boehm:2000gq, Chen:2002yh, Sigurdson:2004zp, Boehm:2004th, Mangano:2006mp, Serra:2009uu, Wilkinson:2013kia, Cyr-Racine:2013fsa, Dvorkin:2013cea, Wilkinson:2014ksa, Escudero:2015yka, Ali-Haimoud:2015pwa, Lesgourgues:2015wza} on DM interactions with baryons, photon, and neutrinos can be directly applied to such a kinetic re-coupling case.
This is because they assume that the interaction rate per Hubble time decreases with the cosmic time and they usually put a constraint on the cross section.
In the CHAMP case, however, the cross section of the Coulomb interaction is ill-defined as well known.
Our study is also complementary with the previous works\,\cite{Dubovsky:2003yn,Dolgov:2013una}, where the mass fraction of electroweak-scale CHAMP is studied and constrained numerically by assuming the interaction rate per Hubble time is well larger than unity throughout the cosmic time.

The organization of this paper is as follows.
In the next section, we calculate the momentum transfer rate, which is  important in the evolution of cosmological perturbations of CHAMP.
In section\,\ref{sec:ttpower}, we follow the co-evolution of cosmological perturbations of CHAMP, baryons, photon, neutrinos, and the gravitational potentials by numerical calculations. We show the resultant $TT/EE$-power spectra in CHAMP models.
We conclude this paper in section\,\ref{sec:concl}.
We adopt the cosmological parameters listed as ``{\it Planck} TT+lowP" in\,\cite{Ade:2015xua} throughout this paper.

\section{Momentum transfer rate}
\label{sec:gamma}
In this paper, we adopt the treatment of the collision term developed in paper I \cite{Binder:2016pnr}.
There it is assumed that the momentum transfer per collision is smaller than the typical DM momentum.
This is validated as long as the CHAMP mass is much larger than any other relevant energy scale.
We can see it in the following way.
In the CHAMP case, the momentum transfer per collision is $\sim T_{b}$ or $\sim \sqrt{T_{b} m_{b}}$ depending on if baryons are relativistic or nonrelativistic.
Here $T_{b}$ is the temperature of baryons (subscript $b$) and $m_{b}$ collectively denotes the mass of baryons.
The typical CHAMP momentum is $\sim \sqrt{T_{\rm Ch} m_{\rm Ch}}$ where $T_{\rm Ch}$ is the temperature of CHAMP and equal to $T_{b}$ as long as the Coulomb scattering is effective.
We consider that the baryon temperature is smaller than $\sim 0.1$\,MeV and the CHAMP mass is $\sim 10^{8}$\,GeV or more.
Thus the momentum transfer per collision is smaller than the typical CHAMP momentum.

The collision term is expanded up to the second order in terms of the momentum transfer.
The Fokker-Planck equation of DM is derived and its perturbation theory is developed in paper I \cite{Binder:2016pnr}.
The overall factor in the collision term of the Fokker-Planck equation, the momentum transfer rate, contains all the microscopic details of collisions.
The expansion is not systematic and includes a certain resummation of higher order terms.
This results in averaging over the momentum transfer in the momentum transfer rate.

The momentum transfer rate is given by 
\begin{eqnarray}
\gamma 
=
\sum_{b = e^{+/-}, p^{+}, {\rm He}^{2+}}
\frac{1}{6 m_{\rm Ch} T_{b}}
\sum_{s_{b}} \int \frac{d^{3} \mathbf{p}_{b}}{(2\pi)^{3}}
f^{\rm eq}_{b} 
(1 \mp f^{\rm eq}_{b})
\int^{0}_{-4\mathbf{p}_{b}^{2}} dt (-t)
\frac{d\sigma}{dt}v
\,,
\end{eqnarray}
where $\sum_{s_{b}}$ is the spin sum of baryons, $(E_{b}, {\bf p}_{b})$ the four momentum of baryons at the local inertia frame, $f^{\rm eq}_{b}$ the thermal distribution per spin of baryons, $d\sigma/dt$ the differential cross section, $v$ the relative velocity, and $t$ the momentum transfer squared (the Mandelstam variable).
Here we take the nonrelativistic limit of CHAMP because we are interested in heavy CHAMP so that its mass is much larger than any other relevant energy scale.
For the Coulomb scattering between CHAMP and baryons, 
the invariant amplitude squared averaged over the initial state spins is given by
\begin{eqnarray}
\label{eq:invampsq}
\overline{
|{\cal M}|^{2}
}
=
256 \pi^{2} \alpha^{2} Z_{b}^{2}
\frac{m_{\rm Ch}^{2} E_{b}^{2}}{t^{2}}
\left(
1 +
\frac{t}{4 E_{b}^{2}}
\right) \,,
\end{eqnarray}
with the fine structure constant ($\alpha \simeq 1/137$) and the electric charge of baryons in units of the proton charge ($Z_{b}$).
With this, the transfer rate is reduced to 
\begin{eqnarray}
\gamma 
=
\sum_{b}
\frac{1}{6 m_{\rm Ch} T_{b}}
8 \pi \alpha^{2} Z_{b}^{2}
\int \frac{d^{3} {\bf p}_{b}}{(2\pi)^{3}}
f^{\rm eq}_{b} 
(1 \mp f^{\rm eq}_{b})
\frac{E_{b}}{|{\bf p}_{b}|}
\int^{0}_{-4p_{b}^{2}} dt
\frac{(-t)}{t^{2}}
\left(
1 +
\frac{t}{4 E_{b}^{2}}
\right)
\,.
\end{eqnarray}

In evaluating the $t$-integral, we encounter a divergence at $t=0$.
This is due to the long-range nature of the Coulomb force.
We regularize the Coulomb divergence by taking into account the Debye screening.
To this end, we replace the upper limit of the integral with the Debye screening scale,
\begin{eqnarray}
k_{D}^{2} 
= 
\sum_{b} 8 \pi \alpha Z_{b}^{2}
\int \frac{d^{3} {\bf p}_{b}}{(2\pi)^{3}}
\left(
\frac{1}{E_{b}}
+ \frac{E_{b}}{{\bf p}_{b}^{2}}
\right) 
f^{\rm eq}_{b}
\,.
\end{eqnarray}
The leading term of the momentum transfer rate is given by
\begin{eqnarray}
\gamma 
\simeq
\sum_{b}
\frac{1}{6 m_{\rm Ch} T_{b}}
8 \pi \alpha^{2} Z_{b}^{2}
\int \frac{d^{3} {\bf p}_{b}}{(2\pi)^{3}}
f^{\rm eq}_{b} 
(1 \mp f^{\rm eq}_{b})
\frac{E_{b}}{|{\bf p}_{b}|}
\ln
\left( 
\frac{4 {\bf p}_{b}^{2}}{k_{\rm D}^{2}}
\right)
\,,
\end{eqnarray}
where the last logarithm is often called the Coulomb logarithm.

\begin{figure}[tb]
 \begin{center}
 \includegraphics[width=0.75\linewidth]{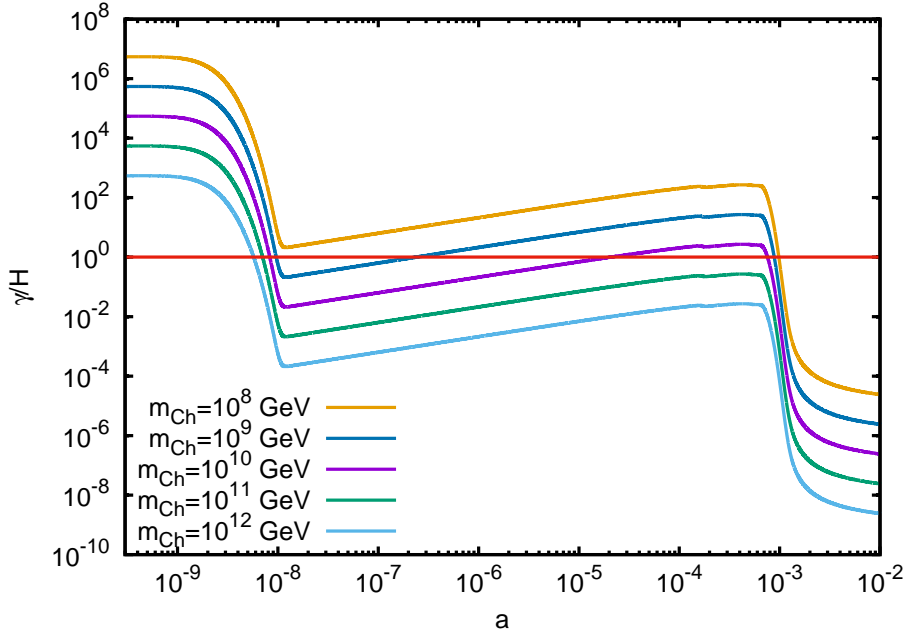}
 \end{center}
 %\vspace{10mm}
 \caption{\sl \small
 Evolution of $\gamma  / H$ with the scale factor ($a$ normalized such that $a=1$ at present). The cosmic time goes from left to right. We take $m_{\rm Ch}=10^{8}, 10^{9}, 10^{10}, 10^{11},$ and $10^{12}$\,GeV. It can be scaled to any CHAMP mass by $\gamma \propto 1/m_{\rm Ch}$ We show the Gamow's criterion ($\gamma/H=1$) for reference. 
 }
\label{fig:CHAMPgam}
\end{figure}

From the above expressions, we can find that before the $e^{+} e^{-}$ annihilation, $e^{+/-}$ make the dominant contributions such that $k_{\rm D} \propto T_{b}$ and $\gamma \propto T_{b}^2/m_{\rm Ch}$.
After the $e^{+} e^{-}$ annihilation, $e^{-}$, $p^{+}$, and He$^{2+}$ make the dominant contribution to $k_{\rm D} \propto \sqrt{Z_{b}^{2} n_{b}/T_{b}}$, where $n_{b}$ collectively denotes the number density of baryons.
Then $p^{+}$ and He$^{2+}$ make the dominant contribution to $\gamma \propto Z_{b}^{2} \sqrt{m_{b}} n_{b} / (T_{b}^{3/2} m_{\rm Ch})$.
In figure\,\ref{fig:CHAMPgam}, we plot numerical results of the evolution of $\gamma / H$ ($H$ is the Hubble expansion rate) as a function of the scale factor ($a$ normalized such that $a=1$ at present).
We can clearly see two sudden drops around $a \sim 10^{-8}$ and $a \sim 10^{-3}$.
The former corresponds to the $e^{+} e^{-}$ annihilation, while the latter to the recombination.
Before the $e^{+} e^{-}$ annihilation, $\gamma / H$ is constant because $\gamma \propto T_{b}^2$ (up to the Coulomb logarithm) and $H \propto T_{b}^2$.
It suddenly drops around the $e^{+} e^{-}$ annihilation and increases slowly with time after that.
This is because $\gamma \propto T_{b}^{3/2}$ (up to the Coulomb logarithm) and $H \propto T_{b}^{2} (T_{b}^{3/2})$ in the radiation-(matter-)dominated era. 
Here we have used $n_{b} \propto T_{b}^{3}$.
Around the recombination, $\gamma / H$ suddenly drops again as the fraction of ionized CHAMP and baryons is reduced.
The small but nonzero fraction of ionized CHAMP and baryons make a tiny contribution to $\gamma / H$ afterwards.

From figure\,\ref{fig:CHAMPgam}, we can see that lighter CHAMP ($m_{\rm Ch}\leq10^{8}$\,GeV) interacts with baryons effectively throughout the cosmic time.
Apparently, these models are not compatible with the observed CMB anisotropies.
This constraint ($m_{\rm Ch} \geq 10^{8}$\,GeV) is comparable with the one from the sea water, but also applicable to decaying CHAMP as long as $\tau_{\rm Ch} \gtrsim 4 \times 10^{5}$\,yr.
Meanwhile, the Coulomb interaction of heavier CHAMP decouples ($\gamma/H$ drops below unity) just after the $e^{+} e^{-}$ annihilation.
Interestingly, it re-couples ($\gamma/H$ increases to about unity) again as the temperature of the Universe decreases.
This is due to the long-range nature of the Coulomb force.
$\overline{
|{\cal M}|^{2}
}$ in eq.(\ref{eq:invampsq}) is larger for a smaller $t$.
The invariant momentum transfer squared ($t$) is roughly $\sim m_{b} T$ after the $e^{+} e^{-}$ annihilation.
Thus $\overline{
|{\cal M}|^{2}
}$ increases as the cosmic temperature cools down.

In figure\,\ref{fig:CHAMPgam}, we show the evolution of $\gamma/H$ only for the region with $a <10^{-2}$.
In fact, since the ionization fraction  rapidly increases 
to unity around the time of reionization, $\gamma / H$  also rises abruptly at $a \sim  a_{\rm reion}$, with 
$a_{\rm reion}$ being the scale factor at the reionization. Depending on the CHAMP mass, 
it can be bigger than unity again.
However, as long as we consider heavy CHAMP, such an increase in  $\gamma / H$ for $a > a_{\rm reion}$ does not affect the CMB anisotropies because, with a heavy CHAMP mass, its number density  is too small to affect the 
reionization optical depth.

%%%%%%%%%%%%%%%%%%%%
%%%%%%%%%%%%%%%%%%%%

Finally let us comment on the Compton scattering between CHAMP and photon.
Our calculation shows that $\gamma / H$ drops below unity when $T \simeq 20 \, {\rm TeV} (m_{\rm Ch} / 10^{8} \, {\rm GeV} )^{3/2} / \epsilon^{2}$.
Here we express the CHAMP charge in units of the proton charge for reference ($\epsilon$).
The Compton scattering decouples in the very early Universe because the cross section is suppressed by $1/m_{\rm Ch}^2$ as mentioned in the previous section.
Therefore in the following we incorporate only the Coulomb scattering between CHAMP and baryons, which is discussed above.

\section{CMB temperature auto-correlation}
\label{sec:ttpower}
Now we follow the evolution of cosmological perturbations of CHAMP, which are governed by (see paper I \cite{Binder:2016pnr})
\begin{eqnarray}
&&
{\dot \delta_{\rm Ch}}
=
-\theta_{\rm Ch}
-\frac{1}{2} {\dot h} \,,
\\
&&
\label{eq:Eulereq}
{\dot \theta_{\rm Ch}}
=
-\frac{{\dot a}}{a} \theta_{\rm Ch}
+k^{2} c_{\rm Ch}^{2} \delta_{\rm Ch}
+\gamma a (\theta_{b} - \theta_{\rm Ch}) \,,
\end{eqnarray}
where a dot denotes the derivative with respect to the conformal time ($\tau$), $\delta_{\rm Ch}$ the density perturbation of CHAMP, $\theta_{\rm Ch} (\theta_{b})$ the bulk velocity of CHAMP (baryon) fluid, and $h$ the gravitational potential in the synchronous gauge\,\cite{Ma:1995ey}.
\footnote{One may wonder how we deal with the residual gauge degrees of freedom in the synchronous gauge because eq.\,(\ref{eq:Eulereq}) does not allow us to fix them by taking $\theta_{\rm Ch} = 0$, which is a conventional choice in the standard CDM model.
We fix the residual gauge degrees of freedom by considering only the physical (adiabatic) mode as an initial condition.
This point will be discussed in detail in the appendix.}
Here we take the Fourier space with the norm of the wavenumber being $k$.
The sound speed squared of CHAMP is given by
\begin{eqnarray}
c_{\rm Ch}^{2} = \frac{T_{\rm Ch}}{m_{\rm Ch}} \left(1- \frac{1}{3} \frac{d\ln T_{\rm Ch}}{d \ln a} \right) \,,
\end{eqnarray}
while the evolution of the CHAMP temperature is described by
\begin{eqnarray}
\frac{d \ln (a^{2} T_{\rm  Ch})}{d \tau}
=
2 \gamma a \left(\frac{T_{b}}{T_{\rm Ch}} -1\right) \,.
\end{eqnarray} 
Hereafter we assume that cosmological perturbations of CHAMP can be well described by the perfect fluid, which is a good approximation as long as the free-streaming length of CHAMP is well smaller than the cosmic scales of interest.
We evaluate the free-streaming scale by the comoving Jeans scale at the matter radiation equality\,\cite{Kamada:2013sh} to find $k_{\rm J} \simeq 2\times10^{6} \,{\rm Mpc} \, (m_{\rm Ch}/10^{8} \, {\rm GeV})^{1/2} (T_{b}/T_{\rm Ch})^{1/2}$, below which we can ignore the free-streaming of CHAMP.

\begin{figure}[tb]
 \begin{center}
 \includegraphics[width=0.75\linewidth]{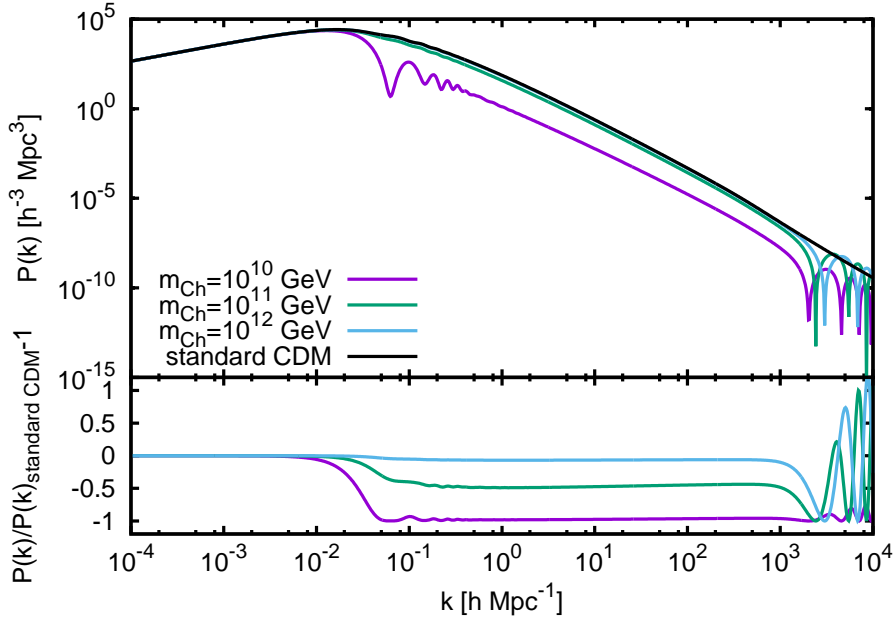}
 \end{center}
  %\vspace{10mm}
 \caption{\sl \small
 Linear matter power spectra at present in the CHAMP models with $m_{\rm Ch}=10^{10}, 10^{11}$, and $10^{12}$\,GeV. 
 In top panel, we also show the spectrum in the standard CDM models, where DM particles do not have any interaction other than the gravitational one. 
 We compare the differences between the CHAMP models and the standard CDM model in the bottom panel. 
 The heavier CHAMP model ($m_{\rm Ch}=10^{12}$\,GeV) is quite close to the standard CDM model at larger length scales ($k \ll 10^3\,h$/Mpc).
 }
\label{fig:CHAMPPk}
\end{figure}

We modify the public code \verb|CAMB|\,\cite{Lewis:1999bs} suitably to follow the co-evolution of cosmological perturbations of CHAMP, baryons, photon, neutrinos, and the gravitational potentials. 
In figure\,\ref{fig:CHAMPPk} (top panel), we show the resultant linear matter power spectra extrapolated to $z=0$.
We can see two oscillatory features of the spectra around $k \sim 10^{3}\,h/$Mpc and $k \sim 10^{-1}\,h/$Mpc.
The former corresponds to the kinetic decoupling of the Coulomb scatterings around the $e^{+} e^{-}$ annihilation ($a \sim 10^{-8}$).
The bottom panel of figure\,\ref{fig:CHAMPPk} shows the differences between CHAMP models and the standard CDM models.
The matter power in CHAMP models at the oscillatory peaks ($k \sim 10^{3}\,h/$Mpc) is larger than the one in the standard CDM.
This is due to the suddenness of the kinetic decoupling as seen in figure \ref{fig:CHAMPgam}.

The cosmological perturbations of CHAMP entering the horizon before the kinetic decoupling start to oscillate with those of photon and baryons.
When CHAMP decouples from baryons suddenly with a maximum bulk velocity of oscillation (when density perturbation takes a zero value), the density perturbations start to be compressed without the photon pressure.
The resultant density perturbations overshoot (become larger than) the oscillation amplitude of density perturbations of photon and baryons. 
Such overshooting, however, does not occur if the kinetic decoupling proceeds more gradually.
We have checked this behavior by setting $\gamma$ by hand to be $d \ln \gamma / d\tau \to d \ln \gamma / d\tau \times 2$ around $\gamma/H=1$.
It is found that the matter power at peaks becomes smaller than the one in the standard CDM and
the overshooting does not occur in a slow decoupling.

The oscillatory feature around $k \sim 10^{-1}\,h/$Mpc is due to the kinetic re-coupling around the matter radiation equality ($a \sim 10^{-4}$), which can be seen in figure \ref{fig:CHAMPgam}.
Afterwards, the subhorizon perturbations undergo damped oscillations around the potential minimum.
While the matter power spectrum in the CHAMP model with $m_{\rm Ch} = 10^{11}$\,GeV shows slight suppression around $k \sim 10^{-1}\,h/$Mpc, the one with $m_{\rm Ch} = 10^{12}$\,GeV is quite close to the matter power spectrum in the standard CDM model except for the oscillatory feature at smaller length scales ($k \sim 10^{3}\,h/$Mpc).
This oscillatory feature disappears if CHAMP is as heavy as $m_{\rm Ch} > 10^{16}$\,GeV.

\begin{figure}[tb]
 \begin{center}
 \includegraphics[width=0.75\linewidth]{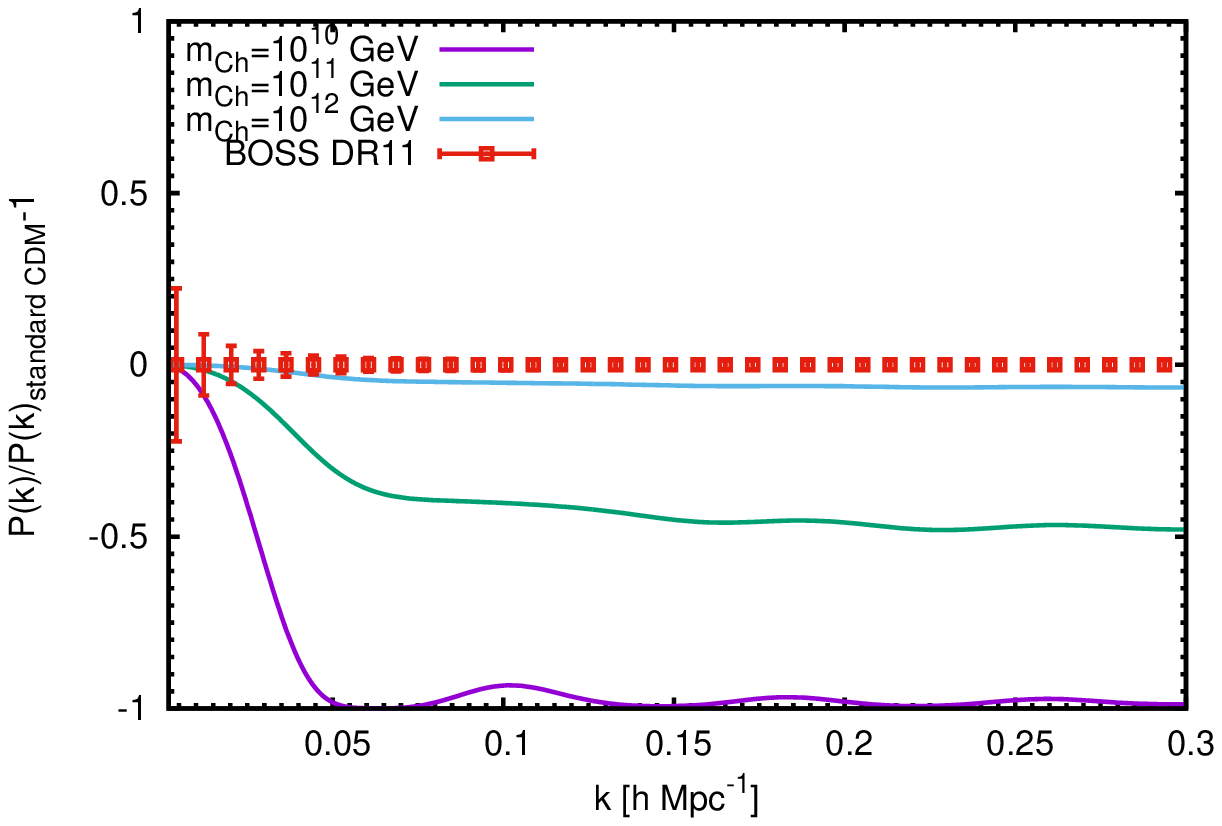}
 \end{center}
  %\vspace{10mm}
 \caption{\sl \small
 The same as in the bottom panel of figure\,\ref{fig:CHAMPPk}, but zooming in on the baryon acoustic oscillations.
 We also show the relative "errors" in the BOSS DR11 data (``post-recon")\,\cite{Anderson:2013zyy} for reference.
 Note that the vertical offset may be compensated by a choice of the template model parameters such as a galaxy bias, which are varied when the template model is fit to the data (see \cite{Anderson:2013zyy} for details).
 }
\label{fig:CHAMPPkdiff}
\end{figure}

Precise measurements of the baryon acoustic oscillations may constrain CHAMP models severely.
To show their potential, in figure\,\ref{fig:CHAMPPkdiff} we compare the differences between CHAMP models and the standard CDM model shown in the bottom panel of figure\,\ref{fig:CHAMPPk} with the relative ``errors" (strictly speaking, the square roots of the diagonal elements of the covariance matrix) of the baryon acoustic oscillations in the clustering of galaxies from the Baryon Oscillation Spectroscopic Survey (BOSS), which is part of the Sloan Digital Sky Survey III\,\cite{Anderson:2013zyy}.
The differences in the CHAMP model with $m_{\rm Ch} = 10^{10}$ and $10^{11}$\,GeV are much larger than the errors.
They appear disfavored by the observations.
Note, however, that a direct comparison between the predicted differences and the observational "errors" should be done with some caution.
This is because we need to introduce additional model parameters such as a galaxy bias to compare the model prediction to the data.
The vertical offset in figure\,\ref{fig:CHAMPPkdiff}, for example, may be easily compensated by a choice of these parameters.
To conclude that the CHAMP model with $m_{\rm Ch} = 10^{12}$\,GeV is disfavored by the observations, we need to perform the fully detailed analysis, which is beyond the scope of this paper.

%%%%%%%%%%%%%%%%%%%%
%%%%%%%%%%%%%%%%%%%%

%%%%%%%%%%%%%%%%%%%%
%%%%%%%%%%%%%%%%%%%%

\begin{figure}[tb]
 \begin{center}
 \includegraphics[width=0.75\linewidth]{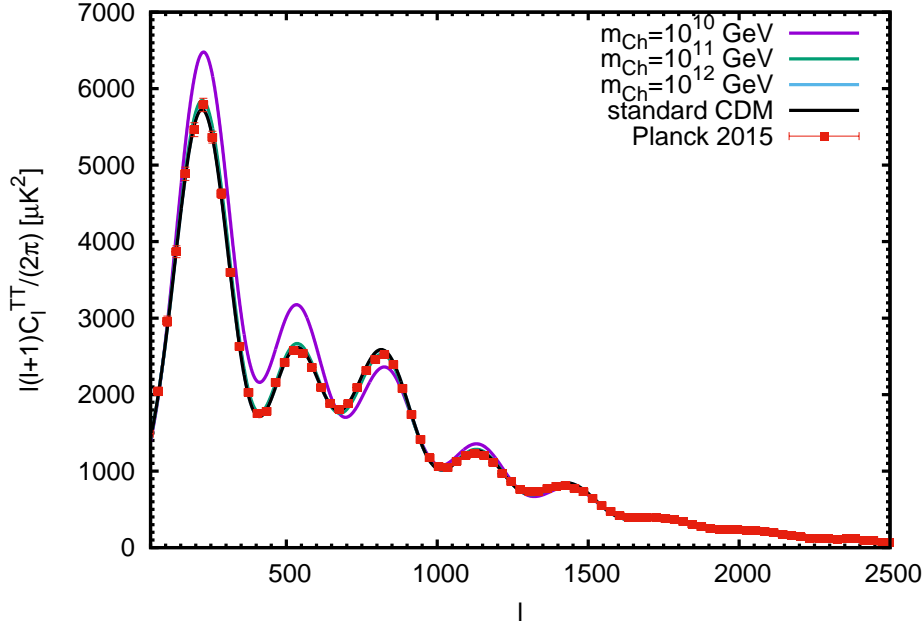}
 \end{center}
  %\vspace{10mm}
 \caption{\sl \small
 CMB $TT$-power spectra in the CHAMP models with $m_{\rm Ch}=10^{10}, 10^{11},$ and $10^{12}$\,GeV. 
 We also show the spectrum in the standard CDM model. 
 Squares with error bars represent the observed spectrum in Planck 2015. 
 The heavier CHAMP models ($m_{\rm Ch}=10^{11}$ and $10^{12}$\,GeV) are quite close to the standard CDM model. 
 We adopt the cosmological parameters listed as ``{\it Planck} TT+lowP" in\,\cite{Ade:2015xua}.
 }
\label{fig:CHAMPTT}
\end{figure}

%\AK{[AK: The following four paragraphs may be too specialized for readers not familiar with CMB anisotropies and too lengthy for experienced readers]}
Next we discuss the effects of CHAMP on the CMB anisotropies.
We utilize the modified version of \verb|CAMB| to predict the $TT/EE$-power spectra in CHAMP models.
We take account of the evolution of $\gamma$ throughout the cosmic time, including its change after the reionization, 
however, as mentioned in the previous section, the rise in $\gamma/H$ for $a > a_{\rm reion}$ does not affect the CMB anisotropies.
In figure\,\ref{fig:CHAMPTT}, we show the resultant $TT$-power spectra.
There are mainly two effects of CHAMP on the $TT$-power spectrum: the enhancement of the even (second, fourth, \dots) peaks and the suppression of the odd (third, fifth, \dots) peaks; the enhancement of the first peak.
The both originate from the decay of the gravitational potential, through different mechanisms.
The mechanisms are known as the baryon drag and the early integrated Sachs-Wolfe (ISW), respectively, in the standard CDM model (see \cite{Hu:1995em, Dodelson:2003ft} for a comprehensive review). 

First let us briefly review these mechanisms in the standard CDM model.
We then describe the differences in CHAMP models.
In the standard CDM model, photon and baryons are compressed along the gravitational potential that is sustained by DM, which is called slow mode in contrast to fast/oscillatory mode, i.e., baryon acoustic oscillation.
The compression leads to the offset of the baryon acoustic oscillation to enhance odd peaks relative to even peaks.
This is called the baryon drag.
Next, let us explain the early ISW effect. 
Suppose that a photon is propagating along the gravitational potential.
The photon gains energy when it moves to the bottom of the gravitational potential, while loses energy when it moves out of the potential.
Though the net energy gain is zero if the gravitational potential is static, the net energy gain is positive if the gravitational potential is decaying.
Actually, the gravitational potential is decaying during the radiation dominated era, while constant during the matter dominated era. 
It may appear that all the photon temperature perturbations that are subhorizon around the matter-radiation equality obtain the positive contribution through the early ISW effect.
Here, however, be reminded that a photon can propagate freely only around and after the recombination, before which the optical depth is huge.
It follows that the early ISW effect is efficient only around the recombination, when the gravitational potential of subhorizon modes is already damped.
The modes just entering horizon around the recombination, i.e., multipoles around the first peak of $TT$-power spectrum, are subject to the early ISW effect.

In the CHAMP case, CHAMP oscillates with baryons through the efficient Coulomb scatterings and do not sustain the gravitational potential unlike the standard CDM. 
This results in a smaller offset of the baryon acoustic oscillation and thus the enhancement of the even peaks and the suppression of the odd peaks in CHAMP models when compared to the standard CDM model.
The decay of the gravitational potential is more drastic and rapid in CHAMP models.
This enhances the early ISW effect, to which the first peak of the $TT$-power spectrum is subject.

\begin{figure}[tb]
 \begin{center}
 \includegraphics[width=0.75\linewidth]{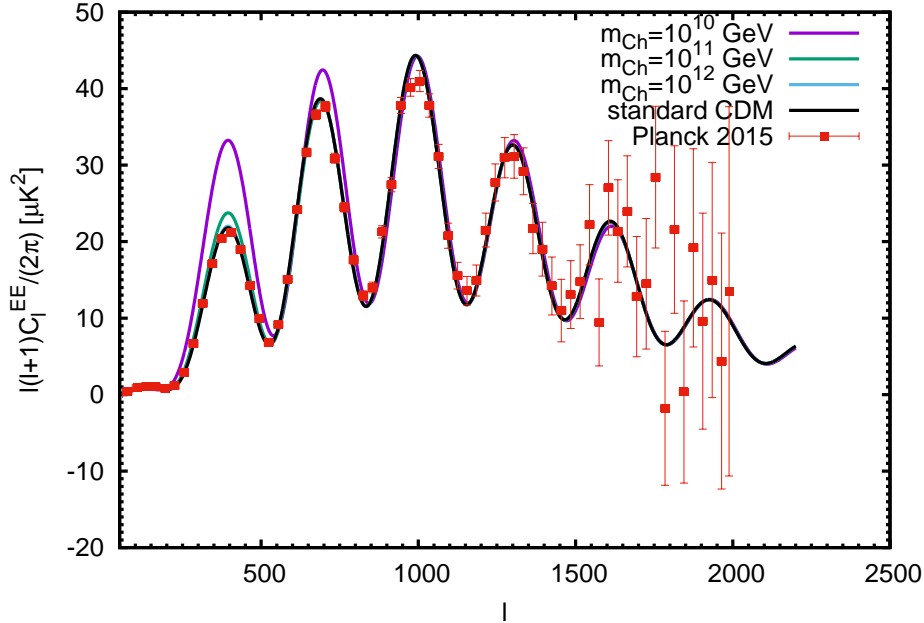}
 \end{center}
  %\vspace{10mm}
 \caption{\sl \small
 The same as in figure\,\ref{fig:CHAMPTT}, but for CMB $EE$-power spectra.
 The heavier CHAMP model ($m_{\rm Ch}=10^{12}$\,GeV) is quite close to the standard CDM model.
 }
\label{fig:CHAMPEE}
\end{figure}

In figure\,\ref{fig:CHAMPEE}, we show the resultant $EE$-power spectra.
%Though intuitive arguments are unfortunately not available for $EE$-power spectra, 
To understand the effects of CHAMP on the $EE$-power spectra, we consider the anisotropic stress (quadrupole) of photon, which are converted to the polarization through the Thomson scattering.
In the leading order of the tight-coupling approximation, it is found that the anisotropic stress is developed around the recombination by the bulk velocity of photon and baryons and the decay of the gravitational potential in the synchronous gauge (specifically see eq.\,(73) in \cite{Ma:1995ey}).
If CHAMP is tightly coupled with baryons, it can be regarded as an increase of ``baryon" mass density.
Thus, in CHAMP models, the baryon mass density dominates the energy density of the photon-baryon plasma earlier than in the standard CDM model, after which the amplitude of baryon acoustic oscillation starts to decay.
Thus the bulk velocity of photon and baryons is smaller in CHAMP models than that in the standard CDM model.
Meanwhile, the decay of the gravitational potential is more drastic in CHAMP models as stressed when discussing the early ISW effect.
Though the two processes cause opposite effects, we checked that the change of the bulk velocity is less important than that induced by the decay of the gravitational potential.
As a result, the $EE$-power spectrum is enhanced in CHAMP models as seen in figure\,\ref{fig:CHAMPEE}.

We can compare the predicted $TT/EE$-power spectra in CHAMP models with the observed ones in the {\it Planck} mission (``Planck 2015")\,\cite{Ade:2015xua}.
The overprediction of the first and the second peaks in the lighter CHAMP model with $m_{\rm Ch}=10^{10}$\,GeV is significant and far beyond the errors in Planck 2015.
The heavier CHAMP model with $m_{\rm Ch}=10^{12}$\,GeV does not show significant deviation from the standard CDM model in the $TT/EE$-power spectrum,
while the middle ($m_{\rm Ch}=10^{11}$\,GeV) leaves a slight enhancement of the first peak only in the $EE$-power spectrum.
If CHAMP is tightly coupled with baryons throughout the cosmic time, all the peaks are affected.
On other hand, in the case of heavy CHAMP, the coupling becomes tight only at a late time, which is called kinetic re-coupling repeatedly in this paper, and only the first peak is affected.
Actually, we need to disentangle the effects of CHAMP and the other cosmological parameters to obtain a robust and precise constraint on the CHAMP mass.
We can, however, infer that the lower bound of the CHAMP mass from the full analysis can be as large as $m_{\rm Ch}>10^{11}$\,GeV.

%%%%%%%%%%%%%%%%%%%%
%%%%%%%%%%%%%%%%%%%%

%%%%%%%%%%%%%%%%%%%%
%%%%%%%%%%%%%%%%%%%%

\begin{figure}[tb]
 \begin{center}
 \includegraphics[width=0.75\linewidth]{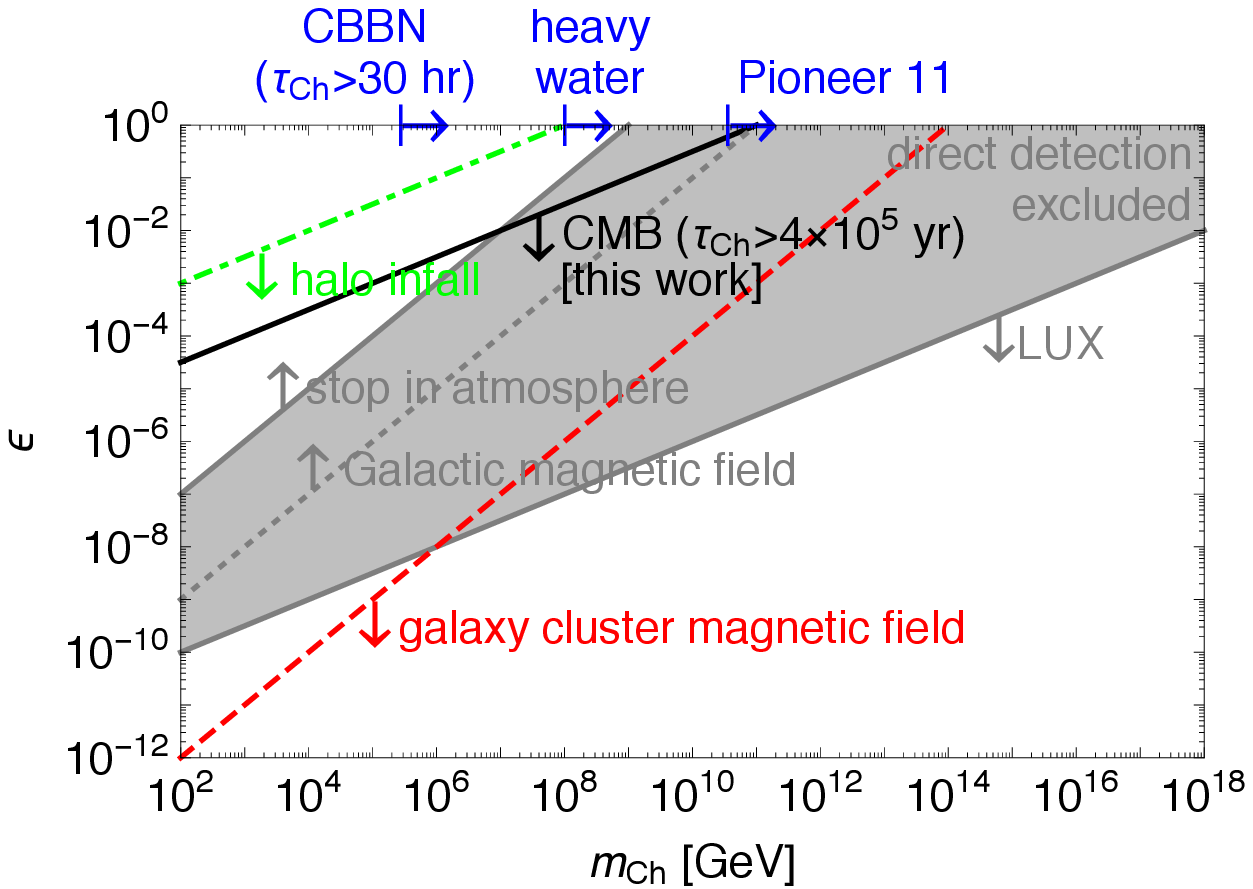}
 \end{center}
  %\vspace{10mm}
 \caption{\sl \small
 Summary of constraints on millicharged particles in the mass-charge plane. 
 The direct detection constraint like the Large Underground Xenon (LUX) experiment\,\cite{Akerib:2013tjd} put an upper bound on the charge, while millicharged particles with a larger charge lose their energy in the atmosphere\,\cite{Dimopoulos:1989hk} and thus are not subject to the constraint.
 In summary, the direct detection constraint excludes only the shaded region.
 The Galactic magnetic field may reduce the number density of millicharged particles in the Galactic disk to open a larger window\,\cite{Chuzhoy:2008zy}.
 It may also relax the severe constraints from a search of heavy isotopes in deep sea water\,\cite{Dimopoulos:1989hk, Agashe:2014kda} and cosmic-ray detectors ({\it Pioneer 11})\,\cite{Dimopoulos:1989hk}. 
 These two constraints and that from the CBBN\,\cite{Pospelov:2006sc, Kohri:2006cn, Steffen:2006hw, Hamaguchi:2007mp, Bird:2007ge, Kawasaki:2007xb, Pradler:2007is, Pospelov:2007js, Kawasaki:2008qe, Jittoh:2008eq, Pospelov:2008ta} are originally given in CHAMP models and their extensions to millicharged particles are nontrivial.
 Hence we show their constraints just on the axis of $\epsilon =1$.
 Millicharged particles lose their energy through the scatterings with electrons in the hot-ionized component of the disk inter-stellar medium and may not maintain a halo\,\cite{Dimopoulos:1989hk}.
 A severe constraint is suggested from the observation that a random walk of millicharged massive particles in the galaxy cluster magnetic field may smear the DM profile\,\cite{Kadota:2016tqq}.
 The constraints are applicable if millicharged particles are (quasi-)stable over the age of the Universe ($\tau_{\rm Ch} \gtrsim 10^{10}$\,yr) unless denoted explicitly.
 The constrains from the CBBN and the CMB anisotropies (this work) are applicable even if millicharged particles are unstable ($\tau_{\rm Ch} \gtrsim 30$\,hr and $4 \times 10^{5}$\,yr, repsectively) but its mass density before the decay is identical to that of DM.
}
\label{fig:summary}
\end{figure}

Before concluding this section, we discuss the implications of our results to millicharged particles.
Millicharged particles are not neutralized at any time as mentioned in section\,\ref{sec:intro}. 
Millicharged particles, however, can not find a charged particle to scatter with as long as its number density is quite small, i.e., its mass is quite heavy with the mass density being fixed.
This is because baryons are neutral after the recombination.
Thus, our results may be applicable to millicharged particles by a simple scaling of $\gamma / H \propto \epsilon^2 / m_{\rm Ch}$.
The bound on the CHAMP mass, $m_{\rm Ch}>10^{11}$\,GeV, could imply that the one for millicharged particles is $m_{\rm Ch} / \epsilon^2 > 10^{11}$\,GeV.
We locate this in the existing constraints and summarize them in figure\,\ref{fig:summary}.

%%%%%%%%%%%%%%%%%%%%
%%%%%%%%%%%%%%%%%%%%

\section{Conclusion}
\label{sec:concl}
We examined the effects of the electric charge of DM on the CMB anisotropies.
Our assumption is that the CHAMP lifetime is longer than the cosmic time at the recombination.
With a help of the recently developed simple treatment of the collision term, we followed the evolution of cosmological perturbations of CHAMP.
A key input in the collision term is the momentum transfer rate of the Coulomb scattering. 
We gave a detailed discussion about how to evaluate and interpret it.
Due to its long-range nature, the scattering rate per Hubble time increases with the cosmic time to lead to a kinetic re-coupling.
By a direct numerical calculation, we followed the co-evolution of cosmological perturbations of CHAMP, baryons, photon, neutrinos, and the gravitational potentials.
We have not assumed that CHAMP and baryons are tightly coupled throughout the cosmic time, while the previous works do.
We found that CHAMP affects the $TT$-power spectrum through the suppression of slow modes and the enhancement of the early ISW effect.
We compared the $TT/EE$-power spectra in CHAMP models with the observed ones in the {\it Planck} mission.
We inferred that the CMB constraint on the mass of electrically charged DM can be as large as $m_{\rm Ch}>10^{11}$\,GeV.
%, while we need to make a detailed analysis to obtain a more precise constraint.

\acknowledgments
This work is partially supported by JSPS KAKENHI Grant Numbers 26247042 (KK), 15K05084 (TT), 25287050, and 25610050 (NY) and MEXT KAKENHI Grant Numbers 15H05889, 16H00877 (KK), and 15H05888 (TT).
NY also acknowledges the financial supports from JST CREST.

%\nocite{*}

\appendix
\section*{Appendix: Residual gauge degrees of freedom in the synchronous gauge}
\label{sec:resgauge}

Some gauge degrees of freedom reside in the synchronous gauge, corresponding to a time-independent shift of the time and the space:
\begin{eqnarray}
\tau \to \tau + \frac{\tilde{\alpha}(\vec{x})}{a} \,, \quad \vec{x} \to \vec{x} + \vec{\nabla} \left(\tilde{\alpha} \int^{\tau} \frac{d\tau'}{a(\tau')} + \tilde{\beta}(\vec{x}) \right) \,.
\end{eqnarray}
Under this gauge transformation the perturbations transform as
\begin{eqnarray}
\label{eq:gaugetrans}
&& h \to h - 6 H \tilde{\alpha} - 2 \Delta \tilde{\alpha} \int^{\tau} \frac{d\tau'}{a(\tau')} - 2 \Delta \tilde{\beta} \,, \quad
\eta \to \eta + H \tilde{\alpha} \,, \notag \\
&& \delta \to \delta - \frac{\tilde{\alpha}}{a} \frac{\dot{\bar{\rho}}}{\bar{\rho}} \,, \quad
\theta \to \theta + \frac{\Delta \tilde{\alpha}}{a} \,.
\end{eqnarray}

From the above expression we can see that the pure gauge modes go like
\begin{eqnarray}
\label{eq:puregaugemodes}
h = A + \frac{B}{(k \tau)^2} \,, \quad \delta = - \frac{1+\omega}{2} \frac{B}{(k \tau)^2}
\end{eqnarray}
outside the horizon in the radiation dominated era.
Here we take the Fourier space and use $\dot{\bar{\rho}} + 3 a H (1+\omega) \bar{\rho} = 0$.
Note that $A = 2 k^2 \tilde{\beta}$ and $B = - 6 H \tilde{\alpha} (k \tau)^2$ are dimensionless constants in time because $H \propto 1/\tau^2$ in the radiation dominated era.

By analyzing the Einstein equations, we can show that there are four independent superhorizon modes in the radiation dominated era\,\cite{Press:1980is, Ma:1995ey}:
\begin{eqnarray}
\label{eq:superhorizonmodes}
h = A + \frac{B}{(k \tau)^2} + C (k \tau)^2 + D (k \tau) \,, \quad \delta_{r} = - \frac{2}{3} \frac{B}{(k \tau)^2} - \frac{2}{3} C (k \tau)^2 - \frac{1}{6} D (k \tau) \,,
\end{eqnarray}
where $\delta_{r}$ is the density perturbation of the radiation that dominates the energy density of the Universe.
As we can see by comparing eqs.\,(\ref{eq:puregaugemodes}) and (\ref{eq:superhorizonmodes}), the $A$- and $B$-modes are just gauge degrees of freedom.
Therefore when we use only the physical $C$- and $D$-modes to set the initial condition, we implicitly exhaust the residual gauge degrees of freedom\,\cite{Press:1980is}.
We drop simply the $D$-mode, because this mode grows more slowly than the $C$-mode. 
Actually the $D$-mode is decaying in the conformal Newtonian gauge\,\cite{Ma:1995ey}, while the $C$-mode is conserved outside the horizon and can be related to the gauge invariant curvature perturbation by $\zeta = - \eta - a H \delta /(\dot{\bar{\rho}}/\bar{\rho}) = - 2C$.

What is the relation with the null bulk velocity of CDM, $\theta_c (\tau) = 0$, found in the literature?
Let us start with the fluid equations of CDM:
\begin{eqnarray}
\label{eq:cdmfluid}
\dot{\delta}_c = - \theta_c - \frac{1}{2} \dot{h} \,, \quad \dot{\theta}_c =  - a H \theta_c \,.
\end{eqnarray}
The superhorizon curvature perturbations, $\delta_c = -1/2 C (k \tau)^2$ and $h = C (k \tau)^2$ in the radiation dominated era, set $\theta_c = \mathcal{O}(C k^3 \tau^2)$ from the equation of continuity (the first equation).
However it does not satisfy the Euler equation (the second equation), which implies $\theta_c \propto 1/ \tau^2$.
Therefore $\theta_c (\tau) = 0$ in the $C$-mode.
In the $B$-mode, on the other hand, it is nonzero.
It can be seen from eq.\,(\ref{eq:gaugetrans}): $\theta = k^2 B / (6 a H \tau^2) \propto 1/a$.
The choice of $\theta_c (\tau) = 0$ is equivalent with that of $B=0$, eliminating one gauge degree of freedom.

How shall we take the initial condition in an interacting dark matter model like a CHAMP model considered in this paper?
The fluid equations of the interacting dark matter are different from eq.\,(\ref{eq:cdmfluid}) by the term like $\gamma a (\theta_r - \theta_i)$ in the right hand side of the Euler equation.
In the tight coupling limit $\gamma / H \gg 1$, $\theta_i = \theta_r$ is an approximate solution of the Euler equation.
As we have seen, the gauge degrees of freedom are exhausted by setting $A=B=0$.
In the $C$-mode $\theta_i = \theta_r = - C (k^4 \tau^3) / 18$, which satisfies the equation of continuity up to the leading order of $k \tau$.

\bibliography{champ_vfinal3}% Produces the bibliography via BibTeX.

\end{document}